\documentclass[12pt,preprint]{aastex}
\usepackage{psfig,rotating,epsfig,graphicx,natbib}

\def\deg   {$^\circ$}

\def\kms{km s$^{-1}$~}

\def\Msun{\mbox{$\rm M_\odot$}}

\shorttitle{Unusual Remote Star Formation} \shortauthors{N. Santiago-Figueroa et al.}
 
\begin{document}
 
\title{A Gaseous Group with Unusual Remote Star Formation}

\author{ N. Santiago-Figueroa\footnote{Department of Physics, Fisk University, 1000 17th Ave N., Nashville, TN 37208; nitza.a.santiago@vanderbilt.edu.} \footnote{Department of Physics and Astronomy, Vanderbilt University, 6301 Stevenson Center, Nashville, TN 37235} \footnote{Department of Astronomy, Columbia University, 550 West 120th Street, New York, NY 10027}, M.E. Putman$^{3}$, J. Werk\footnote{Department of Astronomy and Astrophysics, UCO/Lick Observatory; University of California, 1156 High Street, Santa Cruz, CA 95064}, G.R.\ Meurer\footnote{International Centre for Radio Astronomy Research, The University of Western Australia, M468, 35 Stirling Highway, Crawley, WA 6009, Australia.}, E. Ryan-Weber\footnote{Centre for Astrophysics and Supercomputer, Swinburne University, PO Box 218, Hawthorn, Victoria 3122, Australia} 
\\ \emph{\small Accepted to PASA; August 1 2011} }

\begin{abstract}
 
We present VLA 21-cm observations of the spiral galaxy ESO 481-G017
to determine the nature of remote star formation
traced by an HII region found 43 kpc and $\sim800$ \kms from the galaxy center (in projection).
ESO 481-G017 is found to have a 120 kpc HI disk with a mass of 1.2$\times10^{10}$~${\rm M}_\odot$ and UV GALEX images reveal spiral arms extending into the gaseous disk.  Two dwarf galaxies with HI masses close to $10^8$ ${\rm M}_\odot$ are detected at distances of $\sim$200 kpc from ESO 481-G017 and a HI cloud with a mass of $6 \times 10^7$  ${\rm M}_\odot$ is found near the position and velocity of the remote HII region.   The HII region is somewhat offset from the HI cloud spatially and there is no link to ESO 481-G017 or the dwarf galaxies.  We consider several scenarios for the origin of the cloud and HII region and find the most likely is a dwarf galaxy that is undergoing ram pressure stripping.   The HI mass of the cloud and H$\alpha$ luminosity of the HII region ($10^{38.1}$ erg s$^{-1}$) are consistent with dwarf galaxy properties, and the stripping can trigger the star formation as well as push the gas away from the stars.  

\end{abstract}


\section{Introduction}

Recent observations at both ultraviolet and optical wavelengths have revealed emerging populations of young stars ($<$ 200 Myr) well-beyond the 25$^{th}$ magnitude isophotal radius (R$_{25}$) of a wide variety of galaxies (e.g., Ferguson et al. 1998; Ryan-Weber et al. 2004; Thilker et al. 2007; Werk et al. 2010). Such young populations of stars are seen in the outskirts of spiral galaxies and may signal the ongoing build-up of galaxy disks from the inside-out (White \& Frenk 1991; Thilker et al. 2007).  In other cases, the formation of these outlying stars was clearly triggered by a recent galaxy interaction (Mendes de Oliveira et al. 2004; Boquien et al. 2009).  And in the most distant and luminous cases of so-called outlying star formation, the emission from these young stars may indicate the presence of a star forming dwarf satellite galaxy (Duc \& Mirabel 1994; Boquien et al. 2007).  The origin and fate of these recently-formed stellar populations in the outskirts of galaxies remains an active area of research (Bournaud \& Duc 2006; Mullan et al. 2011).

Deep multi-wavelength observations have the potential to shed light on the nature of outlying star formation.
 In particular, HI synthesis observations show the extent, kinematics, and dynamics of the surrounding gaseous environment. 21-cm observations of systems containing extended, outlying star formation often reveal complex, diverse morphologies, ranging from extended gaseous disks, as in the cases of NGC 2915 (Werk et al. 2010b) and M83 (Huchtmeier \& Bohnenstengel 1981; Bigiel et al. 2010), to very disturbed collisional debris surrounding galaxies (e.g. Boquien et al. 2009). To investigate the origin and potential fates of three outlying HII regions that lie up to 30 kpc from the host galaxy NGC 1533, Werk et al. (2008) use an Australia Telescope Compact Array (ATCA) HI synthesis map of NGC 1533 (Ryan-Weber et al. 2003). These HII regions were found to lie in low-column density, stripped HI gas (1.5 - 2.4 $\times$ 10$^{20}$ cm$^{-2}$) that is offset from the localized HI peaks and has high velocity dispersions ($\sim$ 30 km/s) relative to the rest of the gas.  Where the optical morphology fails to show anything but isolated HII regions, HI observations reveal clues to their formation. In this case of NGC 1533, it appears the outlying HII regions were formed because of an interaction between NGC 1533 and its nearby dwarf companion, IC 2038, that left a large, low-column density HI ring of gas rotating around the galaxy (Ryan-Weber et al. 2004; Werk et al. 2008).

The galaxy of interest for this paper, ESO 481-G017, is an early type spiral galaxy that was observed in a narrow-band H$\alpha$ filter and the R$-$band as part of the Survey for Ionization in Neutral Gas Galaxies (SINGG, Meurer et al. 2006). These observations revealed a high equivalent width H$\alpha$ point-source (emission line dot or ELdot)  that lies 43 kpc (in projection 2.2 times the length of the R-band 25th-magnitude per arcsec$^2$ isophote) from the center of ESO 481-G017. The ELdot associated with ESO 481-G017 (referred to as J0317-22:E1 in Werk et al. 2010) is part of a larger sample of SINGG ELdots with similar observational properties.  The selection criteria of the SINGG ELdots were implemented in an  automated search program to find compact, high-equivalent width sources at projected galactocentric distances greater than 2 $\times$ R$_{25}$ (see Werk et al. 2010 for details). The optical properties of the ELdot associated with ESO 481-G017 are, for the most part, consistent with those of the rest of the sample.  In Werk et al. (2010) we calculated that its H$\alpha$ flux ($9.7\times10^{-16}$ erg s$^{-1}$ cm$^{-2}$) is equivalent to 125 O9V stars, but found no $R$ band flux above the detection limit,
 indicating a recent episode of star formation where comparitively few
 stars existed before.  
  
  Magellan optical spectroscopy has detected H$\beta$, [O III], H$\alpha$, [N II] and [S II] - all typical emission lines of an HII region - from this ELdot; however, the velocity offset between the ELdot and ESO 481-G017 is quite extreme. The spectrum of the HII region reveals that the ELdot has a velocity of 4701 $\pm$ 80 km s$^{-1}$ and ESO 481-G017 has a velocity range 3840-4000 km s$^{-1}$ based on its HI Parkes All-Sky Survey 21-cm emission-line spectrum (HIPASS, Meyer et al. 2004). The large difference in velocity between the main HI disk of the galaxy and these new stars is distinct compared to the other outlying HII regions in Werk et al. (2010) and required further investigation to unravel its origin.   In that light, this paper presents VLA HI observations of ESO 481-G017 and its environment to investigate its relationship (or lack of) to this galaxy.  Understanding the origin of this HII region has implications on the nature of outlying star formation found throughout the universe.

\section{Observations}
 
We observed ESO 481-G017 with the Very Large Array (VLA) telescope for 8 hours in a DnC array configuration at 20 cm and centered on a velocity of 4300 km s$^{-1}$.  The total bandwidth was 6.25 MHz and with 127 channels we obtained 10 km s$^{-1}$ per channel over 1340 km s$^{-1}$.    We used 0317-331 as the flux calibrator and 0312-148 as the phase calibrator.
Data editing, calibration and imaging were completed using the Astronomical Image Processing System (AIPS) 
package of the National Radio Astronomy Observatory (NRAO)\footnote{The National Radio Astronomy Observatory
is a facility of the National Science Foundation operated under cooperative agreement by Associated Universities for
Research in Astronomy.} following the standard procedure for spectral-line observations (Napier et al. 1989).

Calibration of this data set was complicated by the fact that the frequency 1400 MHz for the VLA has substantial interference.  To get rid of the noise we used the task TVFLG in AIPS.  After calibration, images were formed from the \emph{u}-\emph{v} line data by using the AIPS task IMAGR to optimize the beam shape to a synthesized beam of 41$\arcsec$ x 36$\arcsec$ oriented at a
position angle of 86$\degr$. To subtract the continuum we used the task UVLIN and selected clearly line-free channels for the subtraction.   The per channel RMS noise level in the final datacube is typically 1 mJy/beam. 

In addition to the radio data, we employ the SINGG H$\alpha$ and $R$ band
images on this galaxy from Meurer et al.\ (2006), and archival GALEX ultraviolet data
in the FUV and NUV bands.  The $5\sigma$ point source detection limits for the SINGG H$\alpha$ and R-band images are $9.2\times10^{-17}$ erg s$^{-1}$ cm$^{-2}$ and 22.57 ABmag, respectively.
ESO 481-G017 was observed in two over-lapping
GALEX All-sky Imaging Survey (AIS) tiles and we combined these tiles using our
own software.  The total exposure times in the FUV and NUV bands were 320
and 435 seconds, resulting in detection limits of m$_{\rm FUV} = 19.6$ and m$_{\rm NUV} =20.7$ ABmag at a S/N level of 10.  Details on the GALEX satellite and the AIS can be found in Martin et al. (2005).

\section{Results}

\begin{figure}
\begin{center}
\epsscale{1.0} \plotone{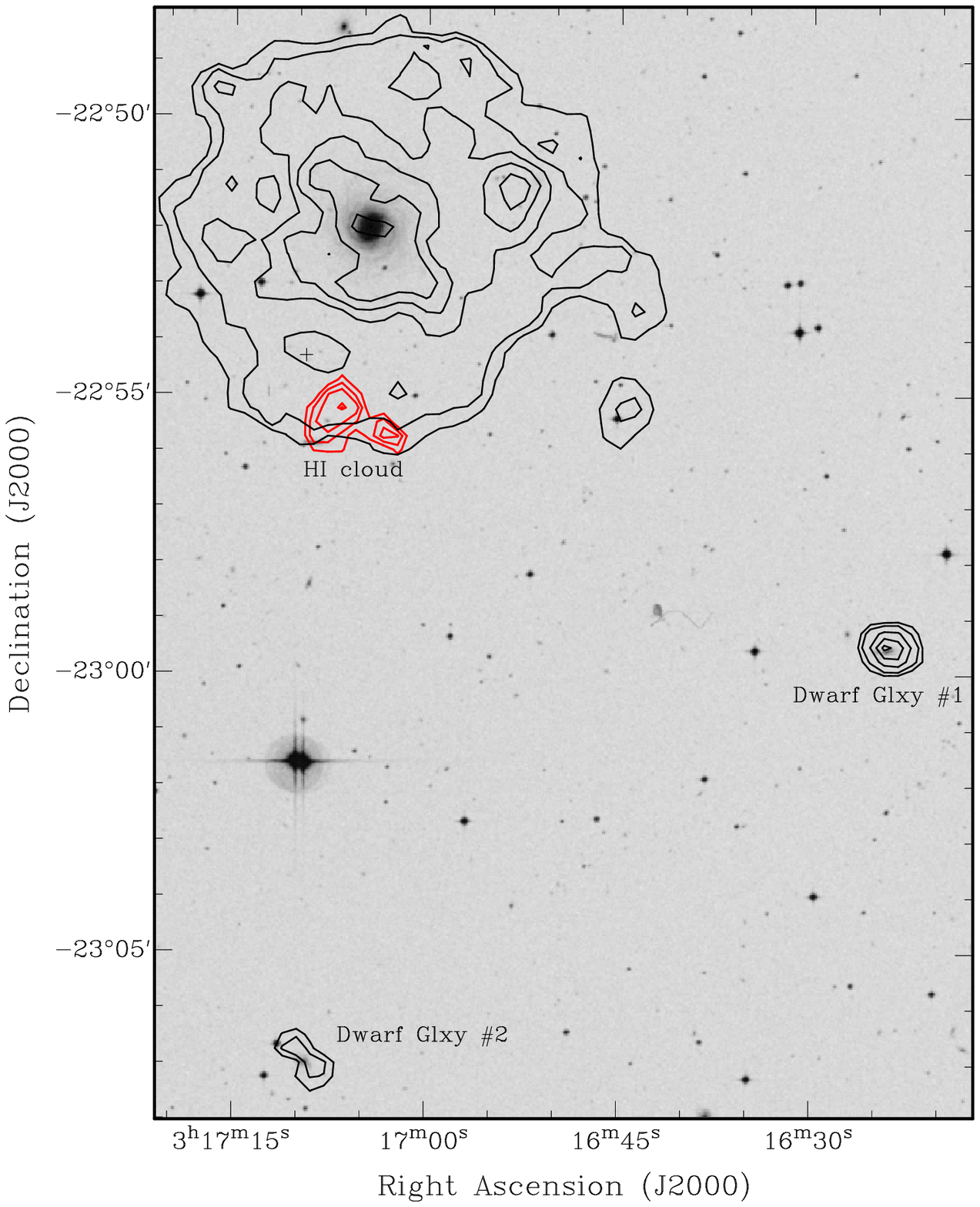} \centering
\caption{ESO 481-G017 and companions with HI contours overlaid at 0.4 ($5\sigma$), 0.8, 1.5, 2.2, 3.0, 3.7, and 4.1 $\times10^{20}$ cm$^{-2}$ on the B-band Digital Sky Survey
(DSS) image. The HI cloud is denoted with red HI contours overlaid at 0.8, 1.6, 2.3, and 3.1 $\times10^{19}$ cm$^{-2}$ and the HII region's position is denoted with a +.  See Table 1 for the positions of the objects.\label{dss}}
\end{center}
\end{figure}

The goal of the observations was to examine the galaxy ESO 481-G017 at 3900 km s$^{-1}$, the HII region at 4700 km s$^{-1}$, and any connection
between these two objects.  We summarize what was found in this section and in Table 1.   Figure~\ref{dss} gives an overview of the VLA field and the detections.  A Hubble
 flow distance of 52 Mpc (for $H_0 = 73\, {\rm km\,s^{-1}\,Mpc^{-1}}$)
 is adopted throughout, as calculated in NED\footnote{NED is operated by the Jet Propulsion Laboratory, California Institute of Technology, under contract with the National Aeronautics and Space Administration.} for ESO 481-G017.

\subsection{ESO 481-G017}

The galaxy ESO 481-G017 is an early type spiral located at
$\alpha_{J2000}$ = 03$^{h}$17$^{m}$04.5$^{s}$, $\delta_{J2000}$ =
-22$\degr$52$\arcmin$00.0$\arcsec$ (see Figures 1 \& 2).  It is
classified ``SB(rs)a pec?'' in the RC3 (de Vaucouleurs et al. 1991) and has an inclination of 18\deg.  Our optical and UV images show the nearly face-on galaxy to have a
high surface brightness bulge transversed by a bar, only apparent
in the R-band.  The bar terminates in an UV and H$\alpha$ bright star
forming ring $\sim$ 3.3 kpc in radius.  Two arms emerge from the
ring, but are only faintly traced by star formation with faint HII regions and UV
knots.  Arm segments can be traced out to a radius of 23.5 kpc to
the NE in the R-band, as well as in very faint diffuse NUV, FUV and H$\alpha$
emission.  This is beyond the R-band 25 ABmag arcsec$^{-2}$
isophotal radius of 19.5 kpc (Werk et al. 2010).  In the B-band Digital Sky Survey
(DSS) images shown in Figure 1, the optical radius is smaller at $\sim 15$ kpc.  Hence, ESO
481-G017 could perhaps be classified as a galaxy with a type-1 extended
UV disk (Thilker et al. 2007), although deeper imaging would be required
for a secure classification.  

In the VLA HI data, ESO 481-G017 extends over a velocity range of 3840-4000 km s$^{-1}$ (Figure 3) and has an HI mass of $1.2 \times 10^{10}$ ${\rm
  M}_\odot$ at 52 Mpc.  This mass was also found by the single dish
observation of Meyer et al. (2004) indicating we are not missing
significant extended emission.  ESO 481-G017 is found to have a large HI
disk with a radius $\sim60$ kpc and also several HI-rich companions that
were previously unknown (see below). The HI disk extends significantly further than the UV and optical described above.
It shows the S-shaped signature of a warp (Figure 3), and a tail of gas to the west with a detached cloud at a
velocity that indicates deeper observations may show a link between this
cloud and the tail.  There is no optical counterpart to this HI cloud.  The stellar mass of ESO 481-G017 can be estimated from
the SINGG R-band absolute magnitude (-21.37 ABmag) resulting in L$_{\rm R} = 2.5 \times 10^{10}$  ${\rm L}_\odot$, the foreground dust corrected B-R=1.35 (Lauberts \& Valentijn 1989), and M/L = 2.5 from Bell et al. (2003), resulting in $6 \times 10^{10}$ \Msun.  
We estimate the dynamical mass for ESO 481-G017 using M$_{\rm dyn} = {\rm V}^{2} {\rm R}/G$, where V is determined from the 50\%
linewidth using an inclination of 18\deg~and R is the HI radius, and find M$_{\rm dyn} =  8.3\times10^{11}$ ${\rm M}_\odot$.   ESO 481-G017's dark matter mass is therefore approximately a factor of 10 larger than its baryonic mass.



\subsection{Dwarf Galaxies}

Two gas-rich companions are found near ESO 481-G017 that can be associated with galaxies whose redshifts were not previously known.  In Figure 1 the HI contours of these detections are shown over the DSS image. We call the two galaxies Dwarf Galaxy \#1 and Dwarf Galaxy \#2 (hereafter, DG \#1 and DG \#2). According to NED, DG \#1 can be identified as APMUKS(BJ) B031411.84-231032.9 and DG \#2 can be identified as APMUKS(BJ) B031457.35-231756.2. DG \#1 is at a distance from the center of ESO 481-G017 of 195 kpc (in projection), has a HI mass of 
$2.3\times10^{8}$ ${\rm M}_\odot$, and a B-band luminosity of $3.8 \times 10^8$ L$_\odot$.  DG \#2 lies at 240 kpc (in projection) from the center of ESO 481-G017, has a HI mass of $9.6\times10^{7}$ ${\rm M}_\odot$, and a B-band luminosity of $3.9 \times 10^8$ L$_\odot$.  The B-band luminosities are estimates taken from the Galactic extinction corrected APM galaxy survey magnitudes (Maddox et al. 1990; Schlegel et al. 1998).
The dwarf galaxies have dynamical masses of 
$3.3\times10^{9}$ ${\rm M}_\odot$ and $9.6\times10^{9}$ ${\rm M}_\odot$ (with no inclination correction),
and cover a velocity range of 4050-4100 km s$^{-1}$ and 4000-4120 km
s$^{-1}$, respectively.  DG \#1 is just
within the frame of the R-band image (see Figure 2), but both galaxies are
beyond the field of view of the H$\alpha$ image. 
DG \#1 and DG \#2 are both visible in the GALEX images.

Approximately 90 kpc (in projection) west of ESO 481-G017 there is a possible low surface brightness (LSB) galaxy companion, located at
$\alpha_{J2000}$ = 03$^{h}$16$^{m}$51.4$^{s}$, $\delta_{J2000}$ = -22$\degr$51$\arcmin$08$\arcsec$ (see Figure 2).  Despite it appearing in the R-band and GALEX NUV image we were not able to identify a unique velocity signature of a dwarf in the HI data, i.e., there is no deviation from the smooth disk gradient.   This includes a close examination of the nearby overdensity of HI seen in the contours of Figure 1.

\subsection{The ELdot and nearby HI Cloud}

In Figure 4 we show the optical H$\alpha$ plus R band image from SINGG near the location of the ELdot (circled). The ELdot was presented in Werk et al. (2010) and confirmed to be an HII region at 4701 $\pm$ 80 km s$^{-1}$ with the detection of multiple emission lines with Magellan spectroscopy.  It has a H$\alpha$ luminosity of $10^{38.1}$ erg s$^{-1}$ (corrected for Galactic extinction (Schlegel et al. 1998)) and is not detected in the R-band SINGG image.  There is a possible 2$\sigma$ detection of the ELdot in the GALEX FUV image (m$_{\rm FUV}=22.4 \pm 0.4$ ABmag).  
A rough estimate of the metallicity of the HII region can be obtained from the [NII]/H$\alpha$ ratio (Pettini \& Pagel 2004).  We find [NII]/H$\alpha$ = 0.2, resulting in 12 + Log O/H = 8.52 and a metallicity of 0.7 solar using the Asplund (2005) solar abundance.   Errors on the Oxygen abundance derived in this manner are estimated at $\pm0.35$ dex.

Although we did not detect HI gas exactly at the position of the isolated HII region, we have discovered a new HI cloud only 11 kpc in projection from the HII region position at a similar velocity (see Figure 4). The HI cloud covers a velocity of 4729-4740 km s$^{-1}$ and the ELdot is at 4701$\pm80$.  On this side of ESO481-G017 the HI disk has a velocity of $\sim3900$ km s$^{-1}$.  The cloud has a HI mass of 6$\times10^{7}$ ${\rm M}_\odot$ at 52 Mpc and, if the linewidth is used to calculate a dynamic mass (potentially not appropriate here), a value of $2.3\times10^{8}$ ${\rm M}_\odot$ is obtained.    There is no obvious optical (or UV) counterpart on the central position of the HI cloud.  There is what appears to be a distant edge-on spiral galaxy with an unknown velocity at the southern edge of the cloud (APMUKS(BJ) B031456.75-230649.7).   Since this galaxy is bright in the UV, but not detected in H$\alpha$, it is highly unlikely to lie at a similar velocity to the cloud and ELdot.

\section{Discussion} 

\subsection{The ESO 481-G017 HI Group}

We find ESO 481-G017 is dominated by a 120 kpc (in diameter) gas disk, although it represents less than 2\% of the galaxy's total dynamical mass within this radius.  Within this gas disk, faint spiral arms are evident in the UV indicating ESO 481-G017 may be an example of a galaxy building itself up from the inside-out.  There are other galaxies with large HI diameters which are comparable to ESO 481-G017 in HI extent, for example:  Malin 1 (220 kpc, Pickering et al. 1997), UGC 1782 (130 kpc, Matthews et al. 2001), and  HIZOA J0836-43 (120 kpc, Donley et al. 2006).   ESO 481-G017 shows some evidence of  disturbance with evidence for a warp in the HI disk and a filamentary extension on the west side.   There is no clear culprit for this disturbance, but the possible LSB galaxy detected in the optical and UV (see Figure 2) could be related.    The HI extension could represent an interaction with a dark matter halo or the accretion of a cold stream of gas.

ESO 481-G017 is found to be part of a small group that contains at least two other gas-rich dwarf galaxies (see Table 1 and Figure 1).  Both of these dwarf galaxies have only approximately 1\% of the total HI mass of ESO 481-G017, but could eventually provide additional gaseous fuel to ESO 481-G017.   There are now a number of sensitive, high resolution HI observations that have detected nearby gas-rich groups and/or HI dwarf companions.  For example, a comparison system is IC 2554 which has a HI mass of 3.2$\times10^{10}$ ${\rm M}_\odot$ and is part of a compact gas-rich group of galaxies (Koribalski et al. 2003).  Another comparison system is that of NGC 2903 which has two gaseous dwarf galaxies nearby with HI masses between 10$^{6-7}$ ${\rm M}_\odot$ (Irwin et al. 2009).   

The fact that the dwarf galaxies detected in HI are at 195 and 240 kpc from ESO 481-G017 in projection is interesting in the context of the location of the gas-rich galaxies in the Local Group.  In the Local Group, within about 270 kpc the vast majority of the dwarf galaxies are undetected or have ambiguous HI detections, while beyond this distance most dwarfs are detected with HI masses $\sim$ $10^{5}$ to $10^{8}$ ${\rm M}_\odot$ (Grcevich \& Putman 2009).  This is most likely due to dwarf galaxies that are closer to the large spiral galaxies being stripped of their HI as they pass through an extended diffuse gaseous halo medium.   Though we do not have deep optical observations to determine if there are nearby gas-free dwarf galaxies, the finding of gas-rich dwarfs only at larger distances could be indicative of a similar phenomenon.

\subsection{What is the ELdot and companion HI cloud?}

The ELdot lies at a projected distance of 43 kpc from the center of ESO 481-G017 and has a H$\alpha$ luminosity of $10^{38.1}$ erg s$^{-1}$ which is equivalent to 125 O9V stars. As mentioned in \S3.3 we have discovered a new HI cloud with no optical counterpart lying 11 kpc (in projection) from the ELdot (see Figures 1 and 4).  The HI cloud has a HI mass of $6\times10^{7}$ ${\rm M}_\odot$ and was found over a velocity range of 4729-4740 km s$^{-1}$, while the ELdot lies at a velocity of 4701$\pm80$ km s$^{-1}$.  The significant offset in velocity of the ELdot and HI cloud from the main galaxy ($\sim$700 km s$^{-1}$) is mysterious, as well as the positional offset of the HI cloud from the ELdot.  Could the HI cloud and ELdot represent leftover material from an interaction with star formation being triggered at the compressed edges of the cloud?  Or do they represent components of a low surface brightness galaxy that has recently begun to form stars again?  In this section we discuss the origin possibilities in the context of its relationship to other known systems.

\subsubsection{The HI Cloud Compared to Other Detached HI Clouds}

Both the ELdot and the HI cloud have a $\sim$700 km s$^{-1}$ velocity offset compared to both ESO 481-G017 and the two new gaseous dwarf galaxies.  This system is unique in that the vast majority of the HI clouds found thus far are closer kinematically to a nearby galaxy.   For example, Kilborn et al. (2006) found 12 HI detections and one extended HI cloud near the NGC 3783 galaxy group. The isolated region of HI gas has a mass of $\sim4\times10^{8}$ ${\rm M}_\odot$ at a distance of 160 kpc from the closest galaxy and it has a velocity offset of $\sim$170 km s$^{-1}$.  The authors think that the origin of this HI cloud is due to tidal debris.  Also, HI observations of the galaxy IC 2554 (Koribalski et al. 2003) detected a large HI plume east of the galaxy as well as two new HI sources with masses similar to our HI cloud.  The two HI sources have a large velocity offset ($\sim200$ and 450 km s$^{-1}$), but still not nearly as large as the velocity offset of our HI cloud.  There are many other detections of HI clouds near galaxies (e.g., Putman et al. 2003;  Oosterloo et al. 2007;  Ryder et al. 2001), but the majority are kinematically (and spatially) in close proximity to the nearby galaxy and most are not shown to be forming stars.

Observations of the Virgo cluster (Kent et al. 2009, Haynes et al. 2007, Oosterloo \& van Gorkom 2005, Minchin et al. 2005; Minchin et al. 2008) have revealed interesting, relatively isolated HI clouds that can be compared to the HI cloud detected here.  The cloud VirgoHI 21 has a HI mass of $\sim$3.4$\times10^{8}$ ${\rm M}_\odot$ and is most likely associated with the harassment of NGC 4254 (Oosterloo \& van Gorkom 2005; Haynes et al. 2007).  It is over 120 kpc away, but both the galaxy and cloud lie at a similar velocity. Simulations of VirgoHI 21 also suggest that this HI cloud was formed from tidal debris (Bekki et al. 2005; Duc \& Bournaud 2008).   The only cloud of similar magnitude in terms of the velocity offset we were able to find is the Virgo Cloud \#1 studied by Kent (2010).  This cloud is offset by 1000 km s$^{-1}$ from the closest galaxy with a velocity.  Similar to other clouds in Virgo, Kent (2010) thinks the cloud's origin can most likely be attributed to ram pressure stripping of the galaxies in that region.   Since ESO481-G017 is not in a cluster environment and this cloud in Virgo is an exception in terms of HI clouds with significant kinematic offsets, it may be unlikely that the HI cloud and ELdot presented here represent the leftovers of galaxy harassment.  Though ESO 481-G017 does have a small filament of HI to the west, it is in the wrong spatial and kinematic direction to have a link to the HI cloud and ELdot.  Gas left behind from the LSB feature seen in the optical and UV in Figure 2 behind ESO 481-G017's HI disk is one possibility, but we do not have a velocity for this feature and it is currently over 70 kpc from the HI cloud and HII region.

\subsubsection{The ELdot in Comparison to Other HII Regions}

With an H$\alpha$ luminosity of $10^{38.1}$ erg s$^{-1}$, the ELdot falls near the peak of model single-burst extragalactic HII region luminosity functions \citep{oey98}. Furthermore, this luminosity places it near the ``saturation luminosity" of a single burst of star formation, signifying that the initial mass function of its stellar population is fully populated.   A lower limit on the stellar mass of the HII region can be obtained by extrapolating a zero-age Salpter IMF from the number of O stars indicated by the H$\alpha$ luminosity \citep[125; see Appendix A of][]{wer08}, resulting in $2 \times 10^4$~\Msun.   The H$\alpha$ luminosity is typical of HII regions in general, however, in comparison with other outlying HII regions (HII regions beyond the optical radius of a galaxy discussed in the introduction), most of which are ionized by a single to a few O stars, it is unusually luminous in H$\alpha$ \cite{wer10}.  For example, the three outlying HII regions of NGC 1533 studied with the High Resolution Channel of $HST$ were found to have H$\alpha$ luminosities between 10$^{37.4}$ and 10$^{37.7}$ erg s$^{-1}$, corresponding to emission from only 20 to 40 O9V stars \citep{wer08}. The outlying HII regions studied as a part of the extended-UV disk of M83 have H$\alpha$ luminosities that range from 10$^{35.5}$ to 10$^{37.4}$ erg s$^{-1}$ \citep{gil07}.  If these lower level outlying HII regions are also present in the ESO~481 - G017 system they would not be detected given the limit of our observations of 10$^{37.7}$ erg s$^{-1}$ at its distance.
The ELdot is most similar in luminosity to the two outlying HII regions of the compact group HCG 16,  with luminosities of 10$^{38.2}$ and $10^{38.5}$ erg s$^{-1}$, and discussed in Ryan-Weber et al.~(2004) and Werk et al.~(2010).  Given that both ESO~481 - G017 (52 Mpc) and HCG 16 (54 Mpc) are considerably more distant than both NGC~1533 (19.1 Mpc) and M83 (4.6 Mpc), we cannot rule out the possibility that these luminous ELdots are unresolved clusters of several HII regions. Along this vein, Werk et al. (2010) note that, given such high luminosities, these ``HII regions'' may not be single isolated star clusters, but rather complexes of young star clusters and/or star formation in a dwarf galaxies.  This latter possibility is discussed in the next section.

\subsubsection{The HI Cloud and ELdot in Comparison to Dwarf Galaxies}

The HI cloud and ELdot may represent the gas and star formation of a dwarf galaxy.    The HI mass of the cloud 
(2.6$\times10^{7}$ ${\rm M}_\odot$) is consistent with the range of HI masses found in dwarf galaxies in the Local Group (Grcevich \& Putman 2009) and the H$\alpha$ luminosity ($10^{38.1}$ erg s$^{-1}$) is also consistent with these dwarf galaxies.  For example, WLM has a H$\alpha$ luminosity of $\sim$2$\times10^{38}$ erg s$^{-1}$ and a HI mass of $6 \times 10^7$ ${\rm M}_\odot$ and Sextans B, Sextans A and IC 1613 have H$\alpha$ luminosities between $10^{37}$ and $10^{38}$ erg s$^{-1}$ and similar HI masses (Mateo 1998).  The ELdot is $<500$ pc in diameter based on the SINGG H$\alpha$ images.  This is somewhat smaller than most dwarf galaxies, but on par with other dwarf companions detected in SINGG images.   The metallicity estimate for the ELdot of 0.7 solar is also consistent with the star formation occurring in gas that has previous hosted stars.

If the ELdot and HI cloud represent a dwarf galaxy, it is possible they are not related to ESO 481-G017; however, one thing that needs to be explained is the offset of the HI cloud and ELdot.  Both the offset and the trigger of star formation traced by the H$\alpha$ could be explained if the dwarf is beginning to pass through an extended diffuse medium, potentially the gaseous halo of ESO 481-G017.   This could result in the gas being stripped from main dark matter and stellar component of the dwarf, as well as star formation being triggered as this gas is compressed during the stripping process.   The high velocity of the dwarf relative to ESO 481-G017 aids in the efficient stripping and explaining the offset of the HI and recent star formation which is seen in the Virgo cluster in galaxies undergoing gas stripping (Crowl et al. 2004; Oosterloo \& van Gorkom 2005).  Given the offset of 11 kpc and the HII region being $<10$ Myr old, the dwarf galaxy needs to be moving at least at $1000$ km s$^{-1}$.   This is roughly consistent with the offset in velocity from ESO 481-G017, but the HII region could also have formed in lower level HI that is not detected here. 

There are potential analogs of offset HI and stellar components in dwarf galaxies in the Local Group.   Two dwarf galaxies in the Grcevich \& Putman (2009) study were listed as ambiguous HI detections because there are nearby HI clouds at similar velocities to the stars, but the clouds do not overlap with the main stellar component.  These two dwarf galaxies (Sculptor and Fornax) are at intermediate distances  (138 and 88 kpc, respectively) and may have had their gas recently stripped by the Milky Way's halo medium. 
 The dwarf galaxy scenario therefore could make sense in terms of what is seen elsewhere in the local universe and may be a complement to the fact that the gas-rich dwarf galaxies are only detected at large radii from ESO 481-G017.  ESO 481-G017 may have an extended diffuse halo medium that can strip dwarf galaxies of their gas and the unusual HI cloud and ELdot may be a consequence of this process.   Deeper optical and HI observations will help to determine if this scenario is correct.

\section{Conclusion}
 
We have discovered that ESO~481-G017 is the brightest member of a small gas-rich group and has a large
extended HI disk ($\sim$60 kpc in radius). There are two gas-rich dwarf companion galaxies detected and a HI cloud at the  
velocity of the HII region, which is 700 km s$^{-1}$ offset from ESO 481-G017 and the new dwarf galaxies.  
The HI cloud and ELdot are offset from each other by 11 kpc.  We investigated several possible origins for the HI cloud and ELdot by considering previously known isolated HI clouds and outlying HII regions.   We find they most likely represent components of a dwarf galaxy and that the offset of the H$\alpha$ and HI may be explained by the dwarf currently undergoing ram pressure stripping (possibly as it moves into ESO 481-G017's extended halo).   This could lead to the trigger of the star formation with the compression of the gas, as well as the gas being moved away from the stars.   It is also consistent with all of the dwarf galaxies  detected in the HI observations (with optical counterparts) being found at large radii ($> 200$ kpc) from ESO 481-G017, as is also found relative to the Milky Way and Andromeda in the Local Group.  

\acknowledgments
Special thanks to Jacqueline van Gorkom for her remarkable collaboration during the data reduction. We also thank students Kathryn Statonik, Maria X. Fernandez, and Jennifer Donovan for their assistance with AIPS. We acknowledge support from the New York NASA Space Grant Consortium, GALEX grant NNX10AF72G and NSF grant AST-0904059.  JW and MEP thank the International Centre for Radio Astronomy Research for hosting them while working on this paper.

\clearpage


\begin{figure}
\begin{center}
\epsscale{1.0} \plotone{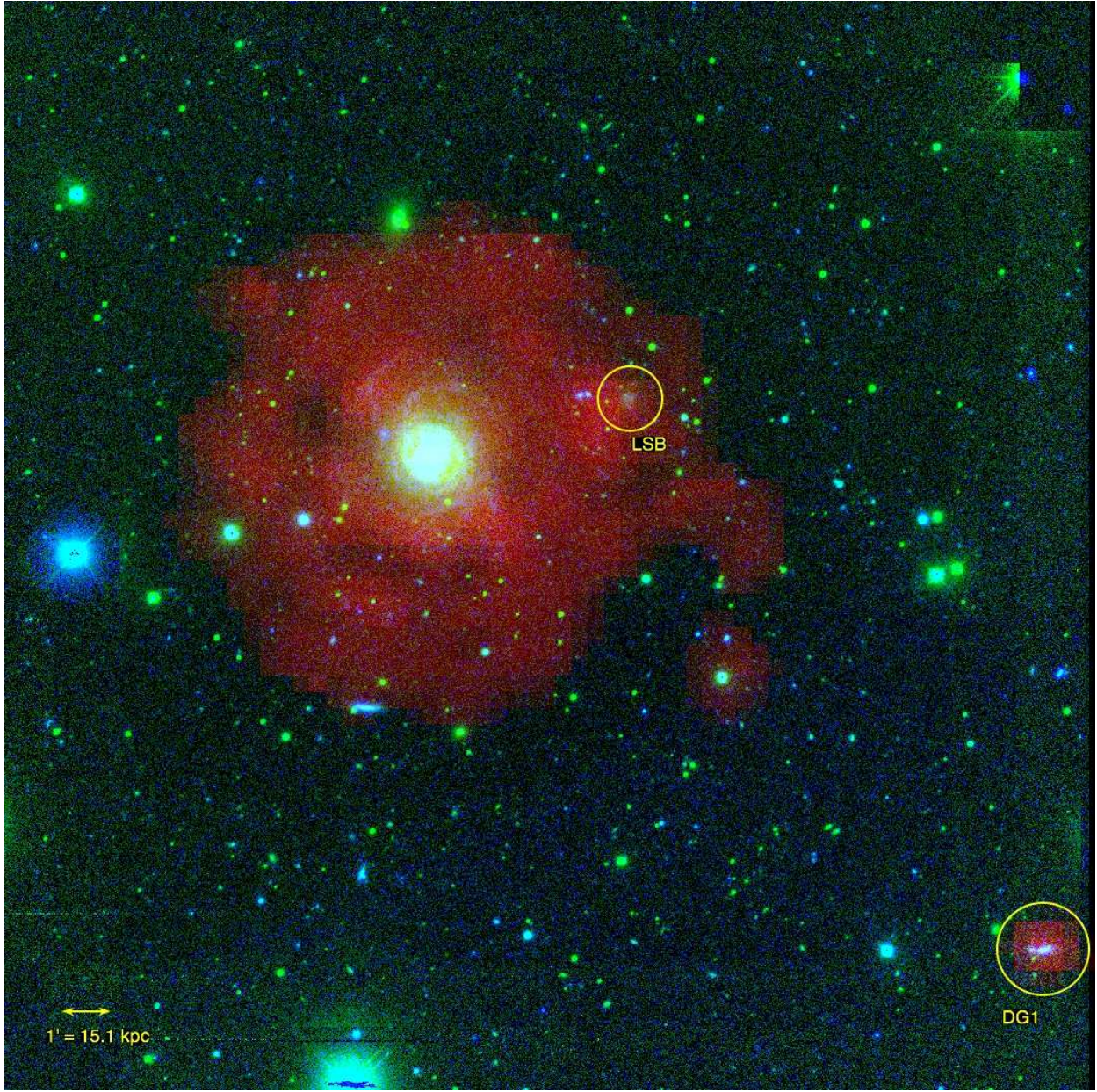}
\caption{The HI (red) over the NUV GALEX image (blue) and R-band image (green) of ESO 481-G017.   The extended nature of the HI is evident, as well as faint spiral arms detected in NUV.  The object labeled Dwarf Galaxy \#1 in Figure 1 is labeled DG1 at the bottom right and a 1\arcmin~scale bar is shown to the bottom left.  The possible LSB galaxy discussed in \S3.2 is also circled.}
\end{center}
\end{figure}

\begin{figure}
\begin{center}
\epsscale{1.0} \plotone{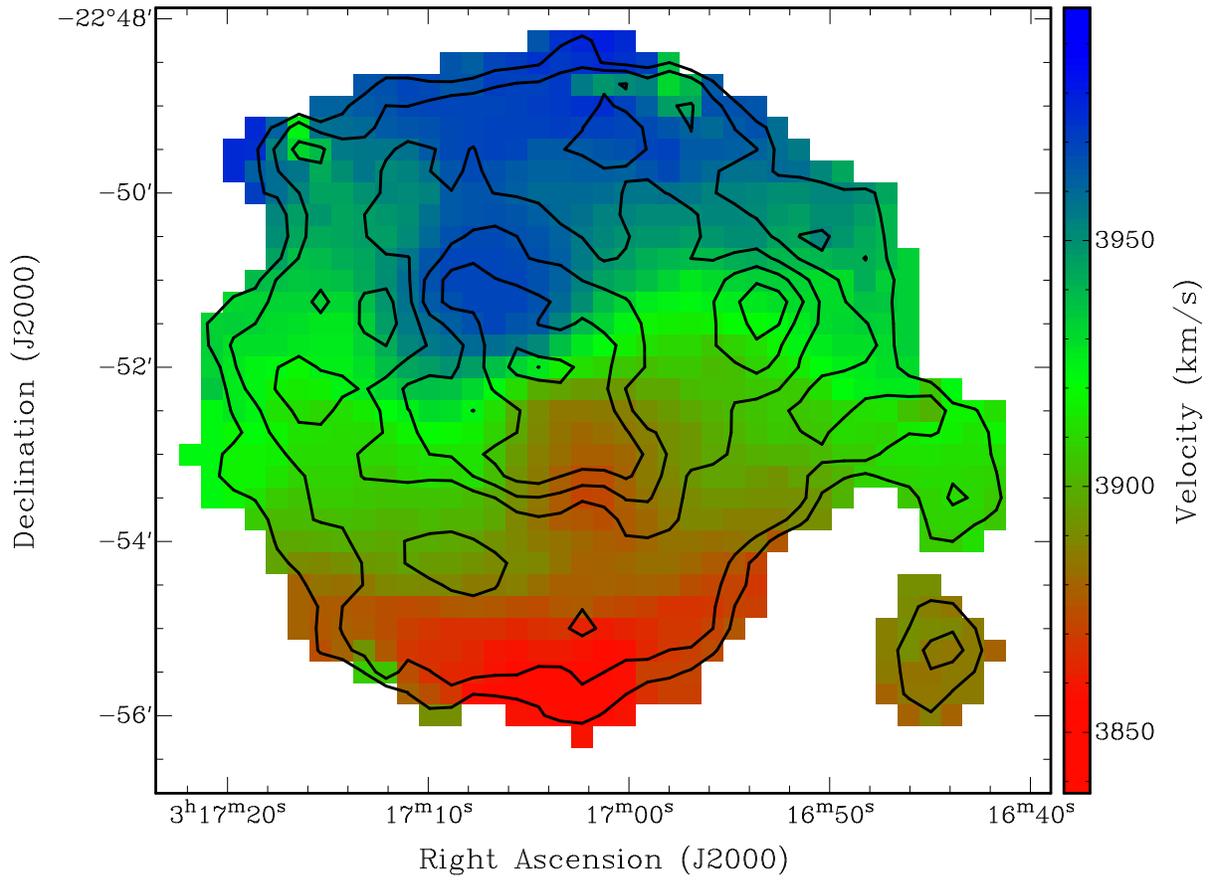} \caption{Velocity moment map of ESO 481-G017 with HI contours from Figure 1 overlaid.}
\end{center}
\end{figure}

\begin{figure}
\begin{center}
\epsscale{1.0} \plotone{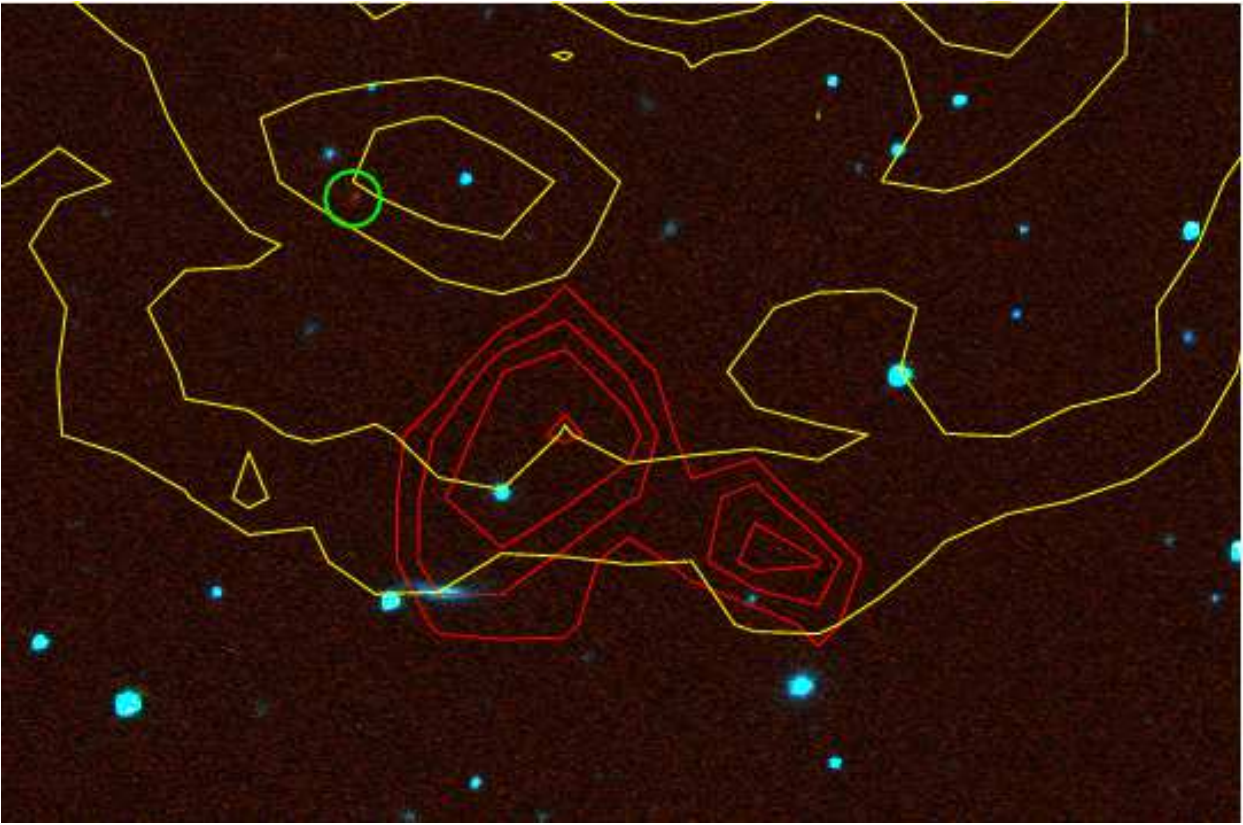} \centering
\caption{Optical H$\alpha$ (red) plus R band (blue) image from SINGG of ESO 481-G017 zoomed in on the area of the HI cloud and HII region (see Figure 1). The HI cloud is denoted with red HI contours, the yellow contours are the HI of ESO 481-G017, and the HII region is circled in green.} 
\end{center}
\end{figure}

\clearpage




\begin{deluxetable}{ccccccc}
  \tabletypesize{\scriptsize}
  \tablewidth{0pt}
  \tablecaption{Properties of ESO 481-G017 and Companions}\label{tab1}
  \tablehead{\colhead{}& \colhead{RA\tablenotemark{a}}& \colhead{Dec\tablenotemark{a}}&\colhead{V$_{\rm HI}$} &\colhead{Distance\tablenotemark{b}}& \colhead{HI Mass}& \colhead{M$_{\rm dyn}$}\\
             \colhead{}&\colhead{}&\colhead{}&\colhead{km s$^{-1}$} &\colhead{kpc} &\colhead{${\rm M}_\odot$}& \colhead{${\rm M}_\odot$} }
  \startdata
 
   ESO 481-G017       & 03:17:04.5   & -22:52:00.0    & 3840-4000   &  -    & $1.2\times10^{10}$ & $1.6\times10^{11}$  \\
   Dwarf Glxy (\#1)   & 03:16:24.3  & -22:59:29.7   & 4050-4100   & 195 & $2.3\times10^{8}$ & $3.3\times10^{9}$\\
   Dwarf Glxy (\#2)   & 03:17:08.8   & -23:07:00.0    & 4000-4120   & 240 & $9.6\times10^{7}$ & $9.6\times10^{9}$\\
   HI Cloud           & 03:17:06.7   & -22:55:30     & 4729-4740   & 58  & $6.0\times10^{7}$ & $2.3\times10^{8}$\\
   HII Region         & 03:17:10    & -22:54:18     & 4700$\pm80$ & 43  &       -           &         -        \\

  \enddata
   \tablenotetext{a} {\scriptsize Approximate central value.}
   \tablenotetext{b} {\scriptsize The distances are in projection from the center of ESO 481-G017 which is at 52 Mpc.}
\end{deluxetable}


\begin{thebibliography}


\bibitem[]{} Asplund, M. 2005, ARA\&A, 43, 481
\bibitem[Bekki et al. 2005]{bek05}Bekki, K., Koribalski, B. S., \& Kilborn, V. A. 2005, \mnras, 363, 21
\bibitem[Bell et al. 2003]{} Bell, E., McIntosh, D.H., Katz, N., Weinberg, M.D. 2003, ApJS, 149, 289
\bibitem[Bigiel et al. 2010]{} Bigiel, F., Leroy, A., Seibert, M., Walter, F., Blitz, L., Thilker, D. \& Madore, B 2010, \apjl, 720, L31
\bibitem[]{} Boquien et al. 2009, \aj, 137, 4561
\bibitem[]{} Boquien et al. 2007, \aap, 467, 93
\bibitem[]{} Bournaud, F. \& Duc, P.-A. 2006, A\&A, 456, 481
\bibitem[Crowl et al. 2005]{cro05} Crowl, H., Kenney, J., van Gorkom, J., Vollmer, B. 2005, \aj, 130, 65
\bibitem[de Vaucoulers et al. 1991]{} de Vaucouleurs, G., de Vaucouleurs, A., Corwin, H. G., Jr., Buta, R. J., Paturel, G., Fouque, P. 1991, Third Reference Catalogue of Bright Galaxies, V. 1-3, Springer-Verlag, New York
\bibitem[Donley et al. 2006]{don06}Donley, J. L., Koribalski, B. S., Staveley-Smith, L., Kraan-Korteweg, R. C., Schr\"{o}der, A., \& Henning, P. A. 2006, \mnras, 369, 1741
\bibitem[]{} Duc, P.-A. \& Bournaud, F. 2008, \apj, 673, 787
\bibitem[]{} Duc, P.-A. \& Mirabel, I.F. 1994, \aap, 289, 83
\bibitem[Ferguson et al. 1998]{fer98}Ferguson, A. M. N., Gallagher, J. S., \& Wyse, R. F. G. 1998, \aj, 116, 673
\bibitem[Gil de Paz et al. 2007]{gil07} Gil de Paz et al. 2007, \apj, 661, 115
\bibitem[Grcevich \& Putman 2009]{jana09}Grcevich J., \& Putman, M. 2009, \apj, 111, 222
\bibitem[Haynes et al. 2007]{hay07}Haynes, M. P., Giovanelli, R., \& Kent, B. K. 2007, \apj, 665, 19
\bibitem[]{} Huchtmeier, W. \& Bohnenstengel, H.-G. 1981, \aap, 100, 72
\bibitem[Irwin et al. 2009]{irw09}Irwin, J. A., Hoffman, G. L., Spekkens, K., Haynes, M. P., Giovanelli, R., Linder, S. M., Catinella, B., Momjian, E., Koribalski, B. S., Davies, J., Brinks, E., de Blok, W. J. G., Putman, M. E., van Driel, W. 2009, \apj, 692, 1447
\bibitem[Kennicutt 1989]{ken89}Kennicutt, R. C. 1989, \apj, 344, 685
\bibitem[Kennicutt 1989]{ken08}Kennicutt, R. C. Jr, Lee, J. C., Funes, J. G., Sakai, S., \& Akiyama, S. 2008, \apj, 178, 247
\bibitem[Kent et al. (2009)]{ken09}Kent, B. R., Spekkens, K., Giovanelli, R., Haynes, M. P., Momjian, E., Cortes, J. R., Hardy, E., \& West, A. A. 2009, \apj, 691, 1595
\bibitem[Kent 2010]{ken10} Kent, B.R. 2010, \apj, 725, 2333
\bibitem[Kilborn et al. 2006]{kil06}Kilborn, V. A., Forbes, D. A., Koribalski, B. S., Brough, S., \& Kern, K. 2006, \mnras, 371, 739
\bibitem[Koribalski et al. 2003]{kor03}Koribalski, B., Gordon, S., \& Jones, K. 2003, \mnras, 339, 1203
\bibitem[]{} Lauberts, A. \& Valentijn, E.A. 1989, The surface photometry catalogue of the ESO-Uppsala galaxies, Garching: European Southern Observatory
\bibitem[McConnachie et al. 2011]{mcc11}McConnachie, A. W., Ferguson, A. M. N., Irwin, M. J., Dubinski, J., Widrow, L. M., Dotter, A., Ibata, R., Lewis, G. F. 2011, \aj, 723, 1038
\bibitem[Martin et al. 2005]{martin+05}Martin, D.C. et al. 2005, \apj, 619, L1
\bibitem[Martinez-Delgado et al. 2010]{mar10}Martinez-Delgado, D., et al. 2010, \aj 140, 962
\bibitem[Mateo 1998]{}Mateo, M. L. 1998, \araa, 36, 435
\bibitem[Matthews et al. 2001]{mat01}Matthews, L. D., van Driel, W., \& Monnier-Ragaigne, D. 2001, \aa, 365, 1
\bibitem[Meurer et al. 2006]{meu06}Meurer, G. R., Gerhardt, O., Hanish, D. J., Ferguson, H. C., Knezek, P. M., Kilborn, V. A., Putman, M. E., Smith, R. C., Koribalski, B., Meyer, M., Oey, M. S., Ryan-Weber, E. V., Zwaan, M. A., Heckman, T. M., Kennicutt, R. C., Lee, J. C., Webster, R. L., Bland-Hawthorn, J., Dopita, M. A., Freeman, K. C., Doyle, M. T., Drinkwater, M. J., Staveley-Smith, L., Werk, J. 2006, \apj, 165, 307
\bibitem[Meyer et al. 2004]{mey04}Meyer, M. J., et al.  2004, \mnras, 350, 1195
\bibitem[Minchin et al. 2005]{min05}Minchin, R., Davies, J., Disney, M., Boyce, P., Garcia, D., Jordan, C., Kilborn, V., Lang, R., Roberts, S., Sabatini, S., \& Driel, van W. 2005, \apj, 622, 21
\bibitem[Minchin et al. 2008]{min08}Minchin, R. F., Davies, M. J., Disney, M. J., Marble, A. R., Impey, C. D., Boyce, P. J., Garcia, D. A., Grossi, M., Jordan, C. A., Lang, R. H., Roberts, S., Sabatini, S., van Driel, W. 2008, arXiv:astro-ph/0508153v1
\bibitem[]{} Mullan, B. et al. 2011, ApJ, 731, 93
\bibitem[Napier et al. 1983]{nap83}Napier, P. J., Thomson, A. R., \& Ekers, R. D. 1983, Proc. IEEE, 71, 1295
\bibitem[Oey et al. 1998]{oey98} Oey, M.S. \& Clarke, C.J. 1998, \aj, 115, 1543
\bibitem[Oosterloo et al. 2005]{oos05}Oosterloo, T., \& van Gorkom, J. 2005, \aa, 457, 19
\bibitem[Oosterloo, Fraternali, Sancisi 2007]{oos07} Oosterloo, T., Fraternali, F. \& Sancisi, R. 2007, \aj, 134, 1019
\bibitem[]{} Pettini, M. \& Pagel, B. 2004, \mnras, 348, 59
\bibitem[Pickering et al. 1997]{pic97}Pickering, T. E., Pickering, Impey, C. D., van Gorkom, J. H., \& Bothun, G. D. 1997, \aj, 114, 1858
\bibitem[Putman et al. 2003]{put03} Putman, M., Staveley-Smith, L., Freeman, K., Gibson, B., Barnes, D. 2003, \apj, 586, 170
\bibitem[Ryan-Weber et al. 2003]{rya03}Ryan-Weber, E. V., Webster, R. L., \& Staveley-Smith, L. 2003, \mnras, 343, 1195
\bibitem[Ryan-Weber et al. 2004]{rya04}Ryan-Weber, E. V., Meurer, G. R., Freeman, K. C., Putman, M. E., Webster, R. L., Drinkwater, M. J., Ferguson, H. C., Hanish, D., Heckman, T. M., Kennicutt, R. C., Kilborn, V. A., Knezek, P. M., Koribalski, B. S., Meyer, M. J., Oey, M. S., Smith, R. C., Staveley-Smith, L., \& Zwaan, M. A. 2004, \aj, 127, 1431
\bibitem[Ryder et al. 2001]{ryd01}Ryder, S. D. et al. 2001, \aj, 555, 232
\bibitem[]{}Schlegel, D., Finkbeiner, D., \& Davis, M. 1998, ApJ, 500, 525
\bibitem[Thilker et al. 2005]{til05}Thilker, D. A. et al. 2005, \apj, 619, 79
\bibitem[Thilker et al. 2007]{til07}Thilker,D. A. et al. 2007, 2007, \apj, 173, 538
\bibitem[Werk et al. 2008]{wer08}Werk, J. K., Putman, M. E., Meurer, G. R., Oey, M. S., Ryan-Weber, E. V., Kennicutt, R. C. Jr., \& Freeman, K. C. 2008, \apj, 678, 888
\bibitem[Werk et al. 2010]{wer10}Werk, J. K., Putman, M. E., Meurer, G. R., Ryan-Weber, E. V., Kehrig, C., Thilker, D. A., Bland-Hawthorn, J., Drinwater, M. J., Kennicutt, R. C., Wong, O. I., Freeman, K. C., Oey, M. S., Dopita, M. A., Doyle, M. T., Ferguson, H. C., Hanish, D. J., Heckman, T. M., Kilborn, V. A., Kim, J. H., Knezek, P. M., Koribalski, B., Meyer, M., Smith, R. C., Zwaan, M. A. 2009, \apj, 139, 279
\bibitem[]{} Werk, J., Putman, M., Meurer, G., Thilker, D., Allen, R., Bland-Hawthorn, J., Freeman, K.C. 2010b, \apj, 715, 656
\bibitem[]{} White, S. \& Frenk, C. 1991, \apj, 379, 52

\end{thebibliography}
\end{document}